\documentclass[twocolumn,showpacs,preprintnumbers,amsmath,amssymb]{revtex4}

\usepackage{caption2}
\usepackage{graphicx}
\usepackage{dcolumn}
\usepackage{bm}
\usepackage{subfigure}
\usepackage{subfigure}
\usepackage{picinpar}
\usepackage{picins}
\begin{document}

\preprint{Phys. Rev. E}

\title{Spatial organization and evolutional period of the epidemic model using cellular automata}

\author{Quan-Xing Liu }
\email{liuqx315@sina.com}
\author{Zhen Jin}%
 \email{jinzhn@263.net}
 \author{Mao-Xing Liu}%
\affiliation{%
Department of mathematics, North University of China,\\
Taiyuan, Shan'xi, 030051, People's Republic of China
}%

\date{\today}

\begin{abstract}
We investigate epidemic models with spatial structure based on the
cellular automata method. The construction of the cellular automata
is from the study by Weimar and Boon about the reaction-diffusion
equations [Phys. Rev. E 49, 1749 (1994)]. Our results  show that the
spatial epidemic models exhibits spontaneous formation of irregular
spiral waves at large scales within the domain of chaos. Moreover
the irregular spiral waves grow stably. The system also shows
spatial period-2 structure at one dimension outside the domain  of
chaos. It is interesting that the spatial period-2 structure will
break and transform to spatial synchronous configuration in the
domain of chaos. Our results confirm that populations embed and
disperse more stably in space than they do in non-spatial
counterparts.
\end{abstract}

\maketitle %
\section{Introduction}

Several theoretical models have shown that population invasion and
dispersion is more stable in space than that in non-spatial
counterparts. More stable mean that a previously unstable
equilibrium point becomes stable under a greater variety of
conditions, or that an equilibrium is approached
faster~\cite{Rohani}. In oscillatory systems where the equilibrium
is a limit cycle or more generally an unstable focus, diffusion or
dispersal will create wave-like patterns. The Lattice Lotka-Voltera
(LLV) model was studied extensively
~\cite{Provata,Tsekouras,Provata2}. In non-linear ecology system,
the two most commonly seen patterns are spiral waves and turbulence.
Spiral waves play an important role in ecological systems. For
example, spatially induced speciation prevents extinction for the
predator-prey models~\cite{Savill,Gurney}. A classical epidemic
model is ordinary differential equations (ODEs), or called
mean-field (MF) approximation~\cite{Murray}. The ODEs methods (or
MF) are based on the assumption that the population is well-mixed,
with the subpopulations involving susceptible, infected, removed,
etc., interacting in proportion to their sizes. Nonspatial theory
typically predicts selection for maximal number of secondary
infectors. Among these epidemic features, the existence of threshold
values is crucial for the spread of an
infection~\cite{Anderson,Anderson2}. A second classical approach
describes spatially extended subpopulations, such as a coupled
map-lattice models~\cite{Dieckmann}, reaction-diffusion equations,
deterministic cellular automata and integrodifference equations
model. In the literature~\cite{Sherratt} the author argues the
numerical simulations in a predator-prey system, and shows that
there are either irregular spatiotemporal oscillations behind the
invasion or regular spatiotemopral oscillations with the form of a
periodic travelling `wake' depending on parameter values.

 In our
paper, the geographic spread of an epidemic can be analyzed as a
reaction-diffusion system, in which both subpolulation exhibit local
random movement, and the algorithm of cellular automata are based on
these paper~\cite{Weimar,Sherratt}. More recently, studies have
shown large-scale spatiotemporal patterns in
measles~\cite{Grenfell_1} and dengue fever~\cite{Cummings}. These
studies have shed new light onto key research issues in spatial
epidemic dynamics, but the detailed theoretical studies are
difficult. The study of population dynamics takes into account the
species distribution in space, interactions between individual
species that are located in the same neighborhood, and mobility of
the various species~\cite{Antal,Droz,Ballegooijen}. These studies
predict the formation of spatial complex structure, phase
transitions, multistability, oscillatory regions, etc. In the
Ref.~\cite{Ballegooijen}, the author studies the
susceptible-infected-resistant-susceptible (SIRS) models with
spatial structure using cellular automata rules, showing the
formation process of the spatial patterns (turbulent waves and
stable spiral waves) in the two-dimensional space and existence of
stable spiral waves in the SIRS model.

 The principal
objectives of the present work is that the
susceptible-exposed-infected-resistant (SEIR) model with spatial
structure is investigated by using cellular automata algorithm. The
SEIR model and its classical ODEs version are presented
in~\cite{Olsen,Grenfell}. In fact, many diseases are seasonal, and
therefore an important question for further studies is how
seasonality can influence spatial epidemic spread and evolution.
Hence, we consider the seasonal parameter, $\beta (t)=\beta_{0}(1 +
\varepsilon \sin(2\pi t)$, where $\varepsilon$ is the fluctuating
amplitude of contact rate. Commonly, we describe the susceptibility,
exposure, infection and recover process in terms of four nonlinear
ODEs. We use $S$ for susceptibles, $E$ for the exposed, $I$ for
infectors, and $R$ for the recovered. The dynamical equations for
SEIR model are
\begin{subequations}\label{eq:2}
\begin{equation}\label{eq:2a}
\frac{dS}{dt}=\mu(1-S)-\beta(t) IS,
\end{equation}
\vspace{-0.5cm}
\begin{equation}\label{eq:2b}
  \frac{dE}{dt}=\beta(t)
IS-(\mu+\delta)E,
\end{equation}
\vspace{-0.5cm}
\begin{equation}\label{eq:2c}
\frac{dI}{dt}=\delta E-(\gamma+\mu)I,\hspace{0.5cm}
\end{equation}
\vspace{-0.5cm}
\begin{equation}\label{eq:2d}
\frac{dR}{dt}=\gamma I-\mu R.\hspace{1.5cm}
\end{equation}
\end{subequations}
Here $\mu$ is the death rate per capacity, $1/\delta$ and $1/\gamma$
are the mean latent and infectious periods of the disease.
$\beta(t)$ is the rate of disease transmission between individuals.
The population can be normalized to $S + E + I + R=1$, so all
dependent variables represent fractions of the population. The
original studies show that the system of \eqref{eq:2} exist three
phase transitions which are the stable behavior, the limit cycle and
the chaotic behavior in the mean-field limit~\cite{Olsen} with
respect to the fluctuating amplitude, $\varepsilon$.

\section{Neighourhood-dependent model}

Generally, studies on the spacial epidemic models show that there
exists spatio-temporal travelling waves ~\cite{Djebali} (e.g. dengue
haemorrhagic fever (DHF)~\cite{Cummings,Vecchio} and
measles~\cite{Grenfell_1}). However, few systems are well enough
documented to detect repeated waves and to explain their interaction
with spatio-temporal variations in population structure and
demography.  The actual epidemic spread is spatio-temporal and local
individual interact. Here we study the individual moving of the
susceptible, exposed, infector and recover, and their diffusion from
one lattice site to another. Then the equations~\eqref{eq:2} read
\begin{subequations}\label{eq:3}
\begin{equation}\label{eq:3a}
\frac{\partial S(\bm{r},t)}{\partial t}=\mu-\beta(t) IS-\mu
    S+D_{1}\nabla^{2}S(\bm{r},t),\hspace{0.3cm}
\end{equation}
\vspace{-0.5cm}
\begin{equation}\label{eq:3b}
   \frac{\partial E(\bm{r},t)}{\partial t}=\beta(t)
   IS-(\mu+\delta)E+D_{2}\nabla^{2}E(\bm{r},t),\hspace{0cm}
\end{equation}
\vspace{-0.5cm}
\begin{equation}\label{eq:3c}
\frac{\partial I(\bm{r},t)}{\partial t}=\delta
E-(\gamma+\mu)I+D_{3}\nabla^{2}I(\bm{r},t),\hspace{1.0cm}
\end{equation}
\vspace{-0.5cm}
\begin{equation}\label{eq:3d}
\frac{\partial R(\bm{r},t)}{\partial t}=\gamma I-\mu
R+D_{4}\nabla^{2}R(\bm{r},t).\hspace{1.2cm}
\end{equation}
\end{subequations}

We study the system~\eqref{eq:3} using cellular automata method,
which is suitable for modeling many reaction-diffusion systems in a
quantitatively correct way based on the Ref.~\cite{Weimar}, and
demonstrate recurrent epidemic spiral waves or traveling waves in an
exhaustive spatio-temporal through the numerical simulation. Simply,
we use $
 \bm c(\bm r,t)$ to denote the vector of individual density in
 position $\bm r$ and at time $t$, and
$\bm{L}(\bm{c}(\bm{r},t))$ to describe the local kinetics; $\bm D$
is the diffusion coefficient matrix. Then the system~\eqref{eq:3}
can be written as
\begin{equation}\label{eq:4}
   \frac{\partial \bm{c}(\bm{r},t)}{\partial
   t}=\bm {L}(\bm{c}(\bm{r},t))+\bm{D}\nabla^{2}\bm{c}(\bm{r},t).
\end{equation}
In the following simulation, we may discard the
equation~\eqref{eq:3d}, since we concern the susceptible, exposed
and infected.

We define this model as follows. Space is made up of a square
lattice of $J\times J$. In each step the individuals randomly move
in its neighborhood. The state of the cellular automata is given by
a regular array of density vector $\bm{c}$ residing on a
two-dimensional lattice. We consider cellular automata with $b_i$
states (denoted by the integers $0, 1, 2, 3, \cdots, b_i$). Here
species state $0$ and $b_i$ are zero population level and maximum
population level respectively. The first step of each time iteration
corresponds to local dynamics, and the state at each spatial lattice
changes independently of the states at other vicinity lattices. The
second part of each time step corresponds to unbiased spatial
movement. The central operation of the cellular automaton consists
of calculating the sum
\begin{equation}\label{eq:5}
    \bm{c}_{i}(\bm{r},t)=\sum_{r'\in N_{i}}
    \bm{c}_{i}(\bm r+{\bm r}').
\end{equation}
Where the summation takes up all of the nearest neighbors of the
cell $\bm{r}$. The neighborhoods can be different for each species
$i$. We use the Moore neighborhood for all $i$ in the
two-dimensional space. i.e.
\begin{eqnarray} \label{eq:6}
   N_{square}=\{(0,0),(1,0),(0,1),(-1,0),(0,-1), \nonumber\\
   (1,1),(-1,1),(1,-1),(-1,-1)\}.
\end{eqnarray}
We normalize values of $\bm{c}_{i}(\bm{r},t)$, and the
$\overline{\bm{c}}_{i}(\bm{r},t)=\bm{c}_{i}(\bm{r},t)/(b_{i}N_{i})$
is local average density of the $\bm{c}_{i}(\bm{r},t)$. The
$\overline{\bm{c}}_{i}(\bm{r},t)$ is always between zero and one.

 From the Ref.~\cite{Weimar}, the
two-dimensional discretization version of Eq.~\eqref{eq:4} takes the
form
\begin{widetext}
\begin{equation}\label{eq:7}
\bm{c}_{i}(\bm{r},t+1)=\bm{c}_{i}(\bm{r},t)+\Delta
t\bm{L}(\bm{c}_{i}(\bm{r},t))+D_{i}
\nabla^{2}\bm{c}_{i}(\bm{r},t),\quad i=1,2,3,4.
\end{equation}
\end{widetext}
Where the \begin{equation}\label{eq:9}
   D_{i}=D_{ii}\frac{\Delta t}{\Delta r^{2}},\quad i=1,2,3,4.
\end{equation} and $D_{i}$ defines the
space scale.

Furthermore  we have
\begin{eqnarray}\label{eq:8}
  \bm{c}_{i}(\bm{r},t+1) &=& {\bm L}^{*}(\overline{\bm{c}}_{i}(\bm{r},t)),\quad i=1,2,3,4.
\end{eqnarray}
Where ${\bm
L}^{*}({\overline{\bm{c}}_{i}(\bm{r},t)})=\overline{\bm{c}}_{i}(\bm{r},t)+
\Delta t {\bm L}({\overline{\bm{c}}_{i}(\bm{r},t)})$.

As $\bm{c}(\bm{r},t)$ is the average output of the CA for the system
\eqref{eq:3}, and therefore it is given by
\begin{eqnarray}\label{eq:9}
c_{j}(\bm r,t+1)=\Big\lfloor b_{j}{ L}^{*}\Big(\frac{ c_i(\bm
r,t)}{b_{i}N_i} \Big) \Big\rfloor +1,
\end{eqnarray}
for species $j$.
 About $\bm{c}(\bm{r},t)$ as output of
the CA described in detail can be found in the paper~\cite{Weimar}.

\section{Numerical results}

We have performed extensive numerical simulations of the described
model, and the qualitatively results are shown here. In cellular
automata simulation, periodic boundary conditions are used and
$\Delta t=0.005$. The space scale $D_{1}=0.2$, $D_{2}=0.05$,
$D_{3}=0.02$ and grid size used in the evolutional simulations is
$100 \times 100$ cells. Every species has 100 states in the
system~\eqref{eq:3}, and more states enable more accurately for
discrete representation of the continuum models, while it is complex
for analysis, and this is described in detail
elsewhere~\cite{Sherratt2}. We have tested that the larger grid size
does not change qualitative result for the evolutional dynamics. The
contact rate fluctuates with the seasons can be approximated in
several ways. Simply we choose sinusoidal force,
$\beta(t)=\beta_{0}(1 + \varepsilon \sin(2\pi t)$, where $0
\leq\varepsilon < 1$, and another more realistic option is term-time
force, which sets transmission rates high during school terms and
low in other place~\cite{Earn}. The spatial patterns evolve from
random initial conditions. The maximum density of susceptible
levels, exposed levels and infected levels are set to 50, 5 and 1,
respectively, in the two-dimensional space. Other initial conditions
have been explored as well, and no change has been observed in the
behavior. In Fig.~\ref{fig:1} and Fig.~\ref{fig:3} four different
snapshots during the temporal evolution of the system are presented
in two-dimensional space. These figures and the following figures
are species density levels as a function of space and time on a gray
scale, with white corresponding to the lowest-density state and
black corresponding to the highest density state.

Fig.~\ref{fig:1} and Fig.~\ref{fig:3} have depicted spatial patterns
in the two-dimensional space under different $\varepsilon$ values
(the fluctuating amplitude of contact rate) respectively. We have
examined the temporal evolution by displaying successive time frame
as a movie, but we are unable to represent this effectively on the
printed page. From evolution snapshots
(Fig.~\ref{fig:1a}-\ref{fig:1d}), one can see that there is no
occurrence of spiral waves (fractal fronts) even if the system
reaches a stable state when the $\varepsilon$ is in certain
interval. The certain interval is turned out to be
$\varepsilon^{*}<\varepsilon<\varepsilon_{\rm c}$ with the parameter
value used in the figure and $\varepsilon^{*}\approx0.048$ and
$\varepsilon_{\rm c}\approx0.305$. Here the critical value
$\varepsilon_{\rm c}\thickapprox0.305$ in the spatial
model~\eqref{eq:3}, which is more than the value of the local
dynamic of the chaotic point of system \eqref{eq:2}~\cite{Olsen}
($\thickapprox0.28$). This result suggests another possible
explanation, which is that populations embed and disperse more
stably in space than they do in non-spatial
counterparts~\cite{Hassell1,Rohani2}. As the $\varepsilon$
($\varepsilon>\varepsilon_{\rm c}$, in the domain of chaos) and time
increase, the dynamical patterns with
 fractal fronts (spirals waves) of spatial structures occur, and become larger and more stable (see the
Fig.~\ref{fig:3}). The CA models are generally based on qualitative
rather than quantitative information about the system. It is hardly
to detect the density calculated over the entire lattice when the
system is enough larger. Hence, in order to investigate
quantitatively the evolution of the system \eqref{eq:3}, we give the
results by one-dimensional space. An explicit visualization of
spatial organization within the lattice is provided by the
space-time plot of Fig.~\ref{fig:2} and Fig.~\ref{fig:4}.

\begin{minipage}[c]{0.45\textwidth}
 \subfigure[ 1 step]{%
  \label{fig:1a}
  \hspace{-0cm}\scalebox{0.40}{\includegraphics*[170,292][438,560]{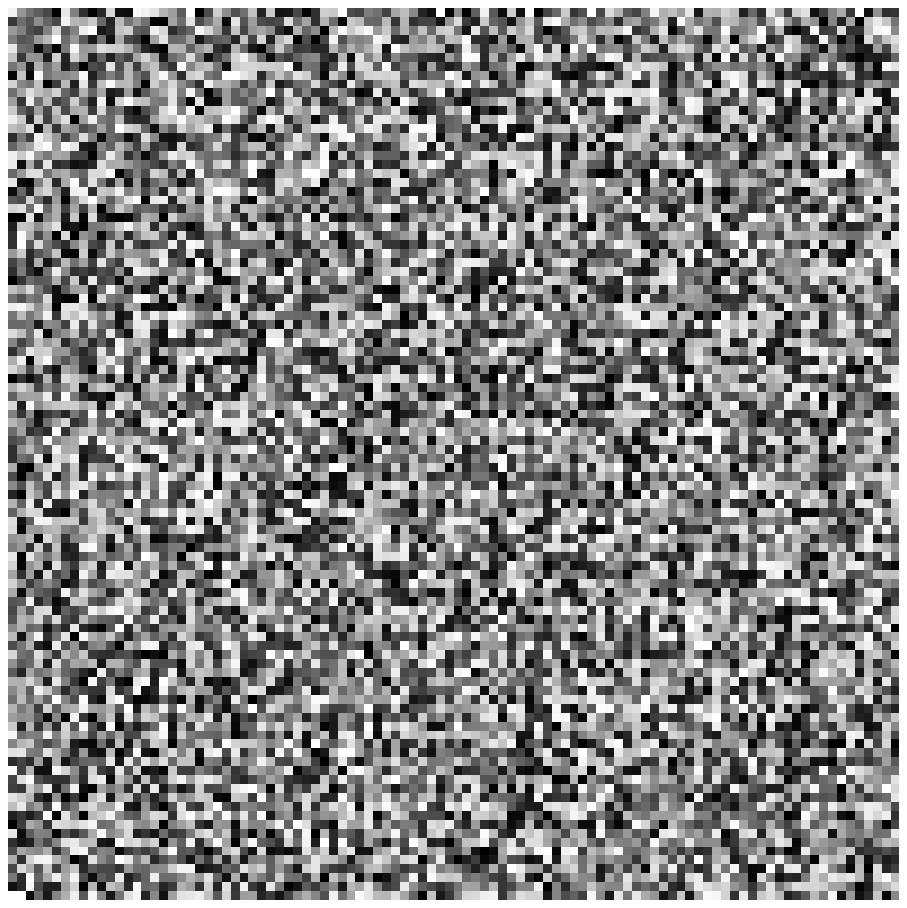}}}
  \subfigure[ 5250 steps]{%
  \label{fig:1b}
  \scalebox{0.40}{\includegraphics*[170,292][438,560]{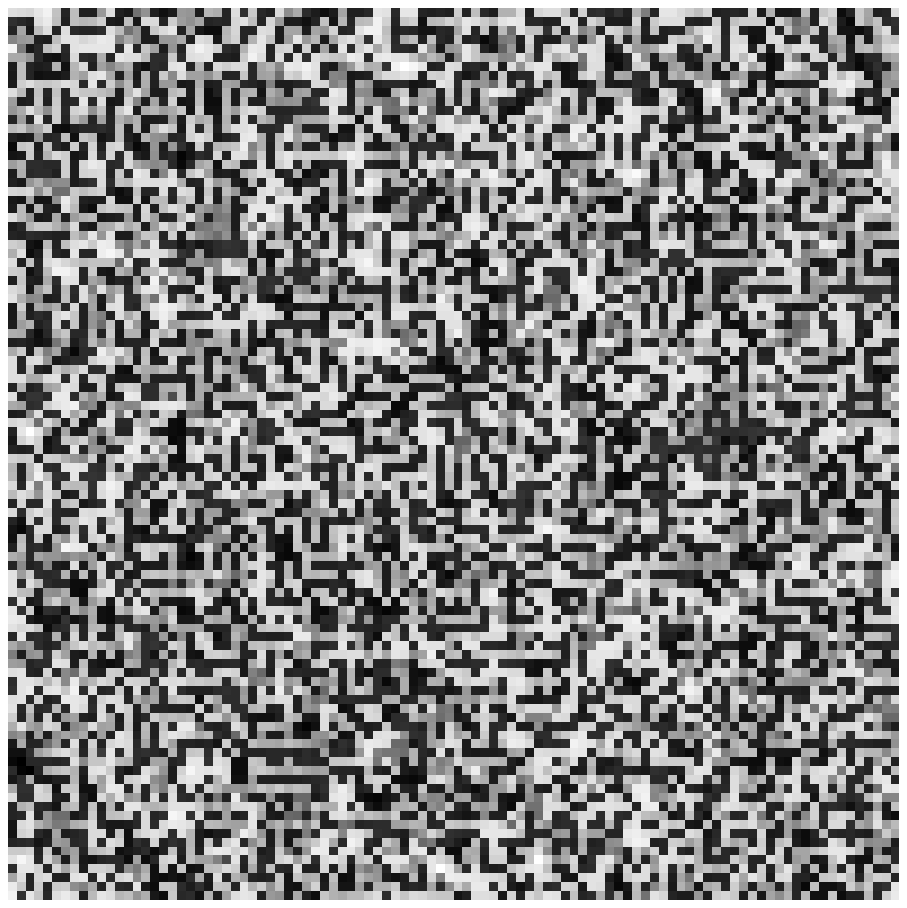}}}\\
   \subfigure[ 5300 steps]{%
  \label{fig:1c}
 \hspace{-0cm}\scalebox{0.40}{\includegraphics*[170,292][438,560]{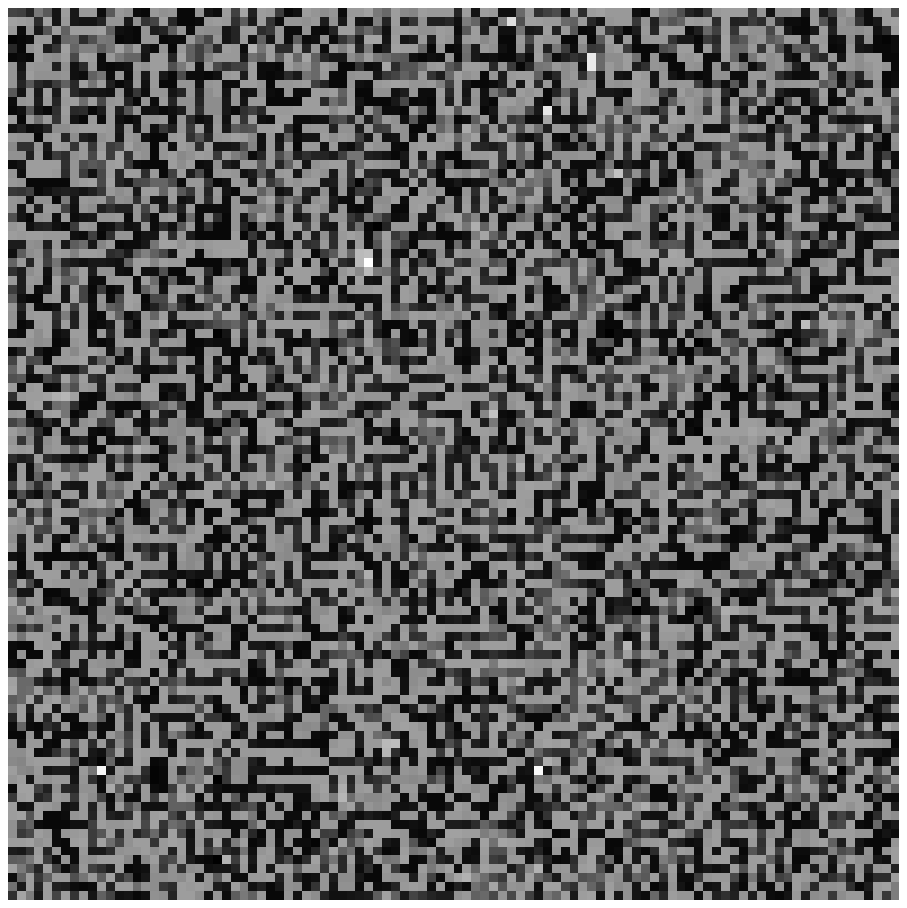}}}
  \subfigure[ 5350 steps]{%
  \label{fig:1d}
  \scalebox{0.40}{\includegraphics*[170,292][438,560]{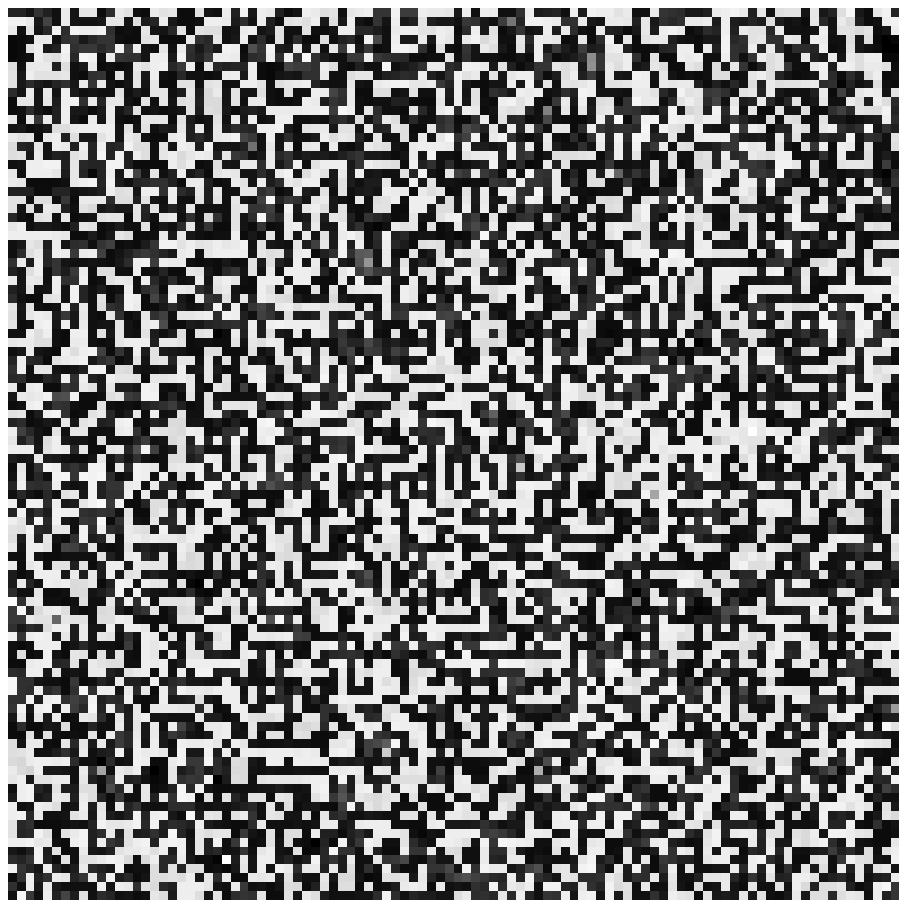}}}
  \caption{A typical simulation shows four snapshots of the evolution in
  two-dimensional space with the parameters $\mu=0.02$, $\delta=35.84$,
  $\gamma=100$, $\beta_{0}=1800$ and $\varepsilon=0.23$. The figures plot
  susceptible density levels as a space on grey scale. The exposed and infected distribution has a qualitatively similar form.}\label{fig:1}
  \vspace{0.5cm}
\end{minipage}

In Fig.~\ref{fig:2a} the three time series displaying of the density
of susceptible, exposed and infected for the first 10000 steps by CA
are given. Self-sustained oscillations of the three time series
develop (see the Fig.~\ref{fig:2a}). The amplitude of oscillations
increase with the increase of fluctuating amplitude of the infection
rate, $\varepsilon$ (compared the Fig.~\ref{fig:2a} and
\ref{fig:4a}). In fact, large oscillations will lead to stochastic
extinction of the species, when the value of fluctuating amplitude
is more than $\varepsilon_{\rm c}$ in the domain of chaos. In
Fig.\ref{fig:2b}-\ref{fig:2c}, spatio-temporal pictures of the
susceptible and exposed are plotted respectively, where time
increases from bottom to top and the horizontal axis represents the
spatial location. From the Fig.\ref{fig:2b}-\ref{fig:2c}, it is
clearly seen that the whole system shows the spatial period-2
structure when the $\varepsilon$ is between $\varepsilon^{*}$ and
$\varepsilon_{\rm c}$ (later we will give the case when
$\varepsilon$ is smaller than a critical value $\varepsilon^{*}$).

\begin{minipage}[c]{0.45\textwidth}
   \subfigure[]{%
  \label{fig:2a}
  \scalebox{0.28}{\includegraphics*[90,262][480,560]{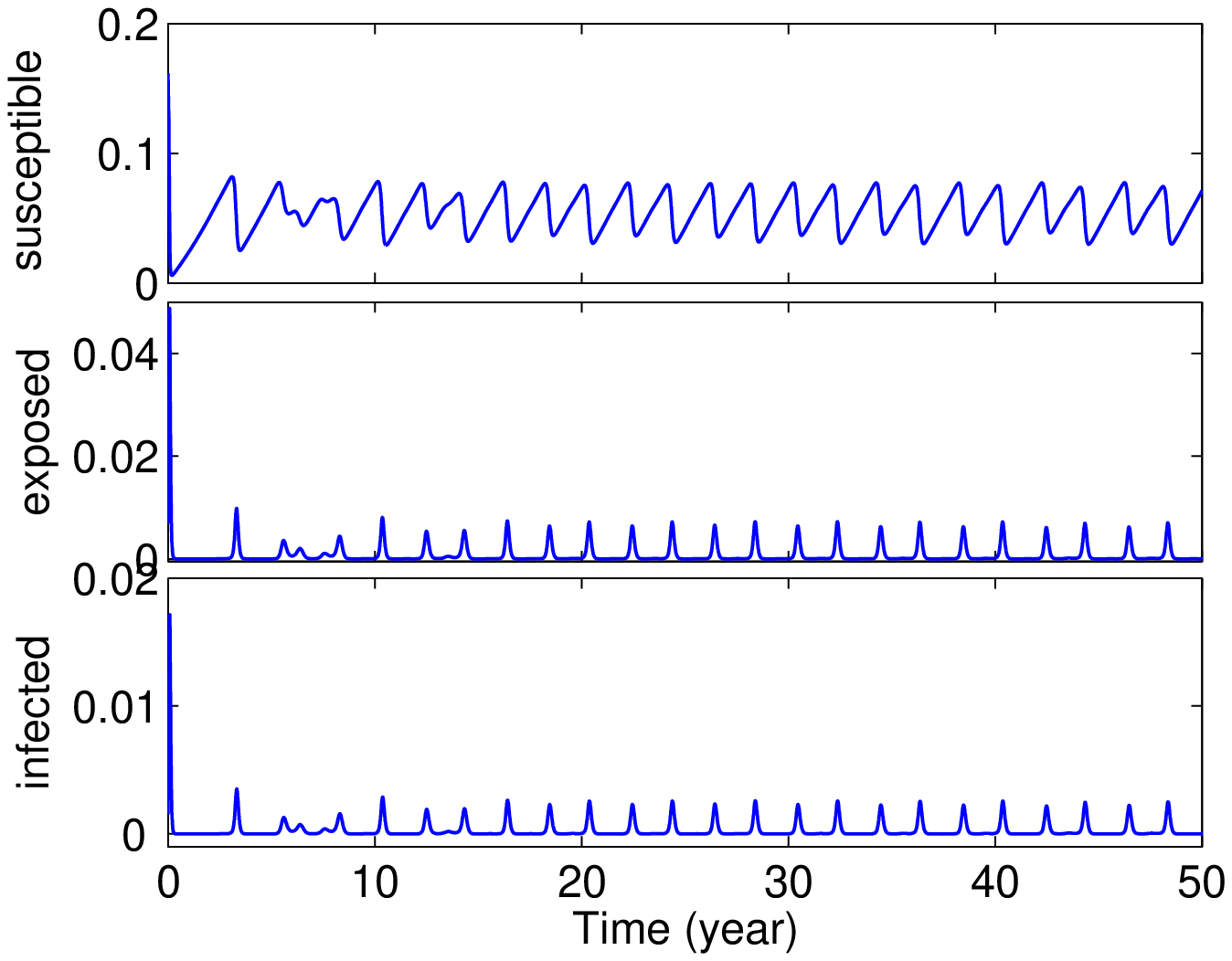}}}\hfill
  \subfigure[]{%
  \label{fig:2b}
  \scalebox{0.28}{\includegraphics*[90,262][480,560]{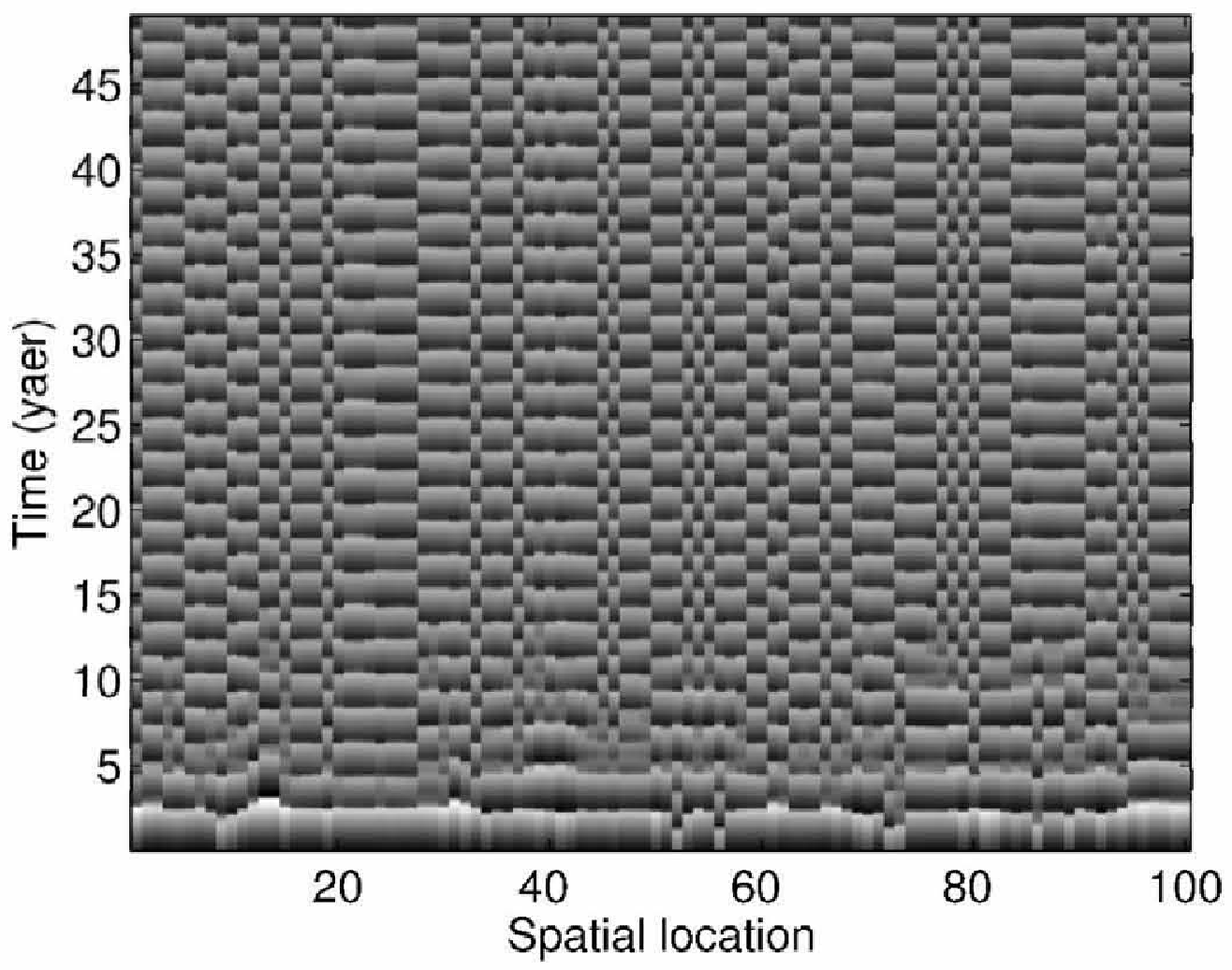}}}\\
  \subfigure[]{%
  \label{fig:2c}
   \scalebox{0.28}{\includegraphics*[90,262][480,560]{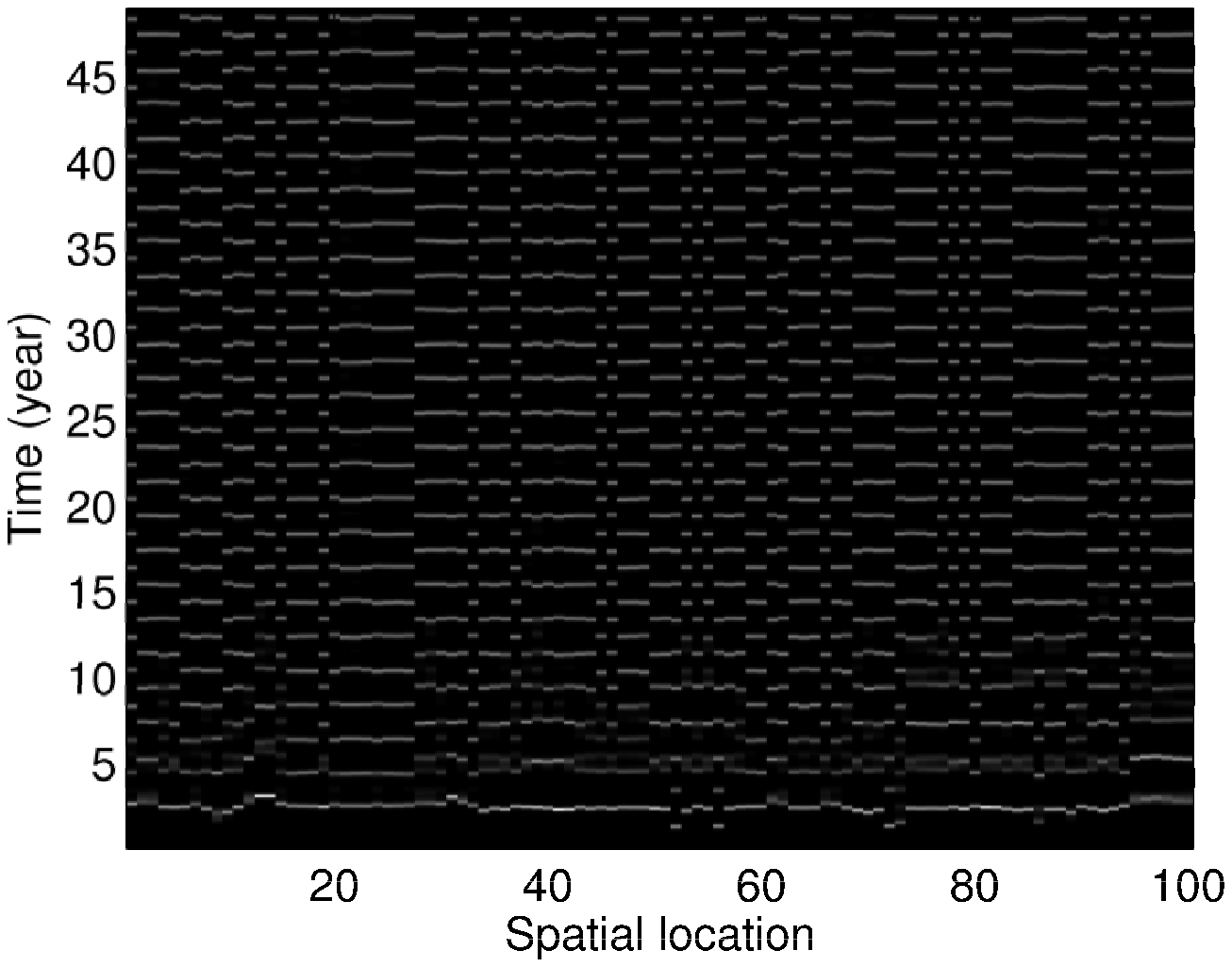}}}\hfill
  \subfigure[]{%
  \label{fig:2d}
  \includegraphics[width=3.5cm]{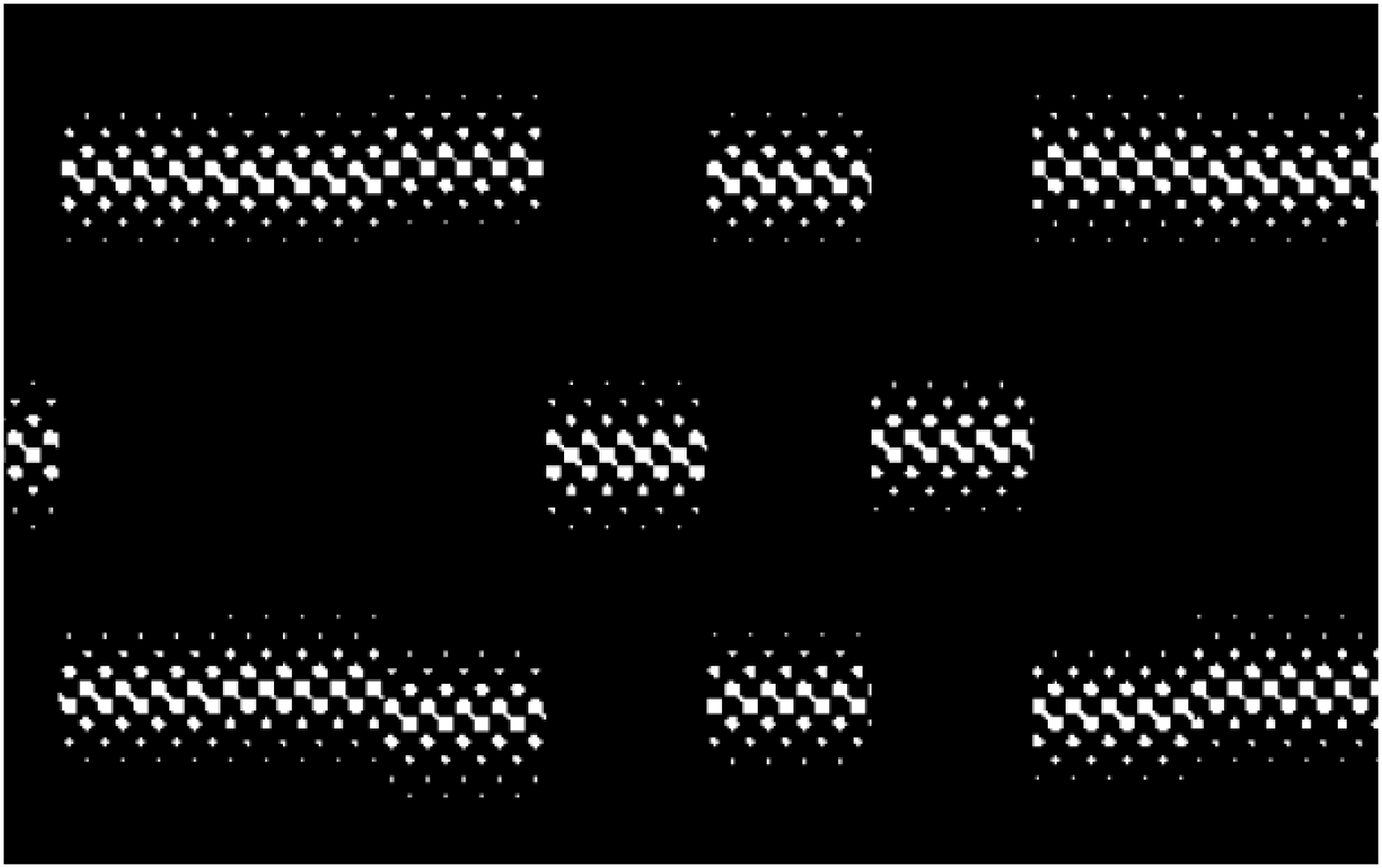}}
   \caption{The spatial period-2 structure results for the system \eqref{eq:3}
  in one-dimensional space in cellular automaton models and the parameters are
  the same as those in Fig.~\ref{fig:1}. We use a spatial domain of 100 lattices;
  10000 successive time iterations are plotted. In (a) we show  the species density as a
  function of time. The figure (b) and (c) show the susceptible and exposed
  density levels as a function of space and time on a grey scale respectively.
  The behavior of the infected is qualitatively similar. At this scale the spatial discretization in
  not really visible, and therefore we have enlarged one region of the plane in figure (c) as figure (d), in order
  to illustrate this discretization.}\label{fig:2}
\vspace{0.5cm}
\end{minipage}

\begin{minipage}[c]{0.45\textwidth}
 \subfigure[1 step]{%
  \label{fig:3a}
  \hspace{-0cm}\scalebox{0.40}{\includegraphics*[170,292][438,560]{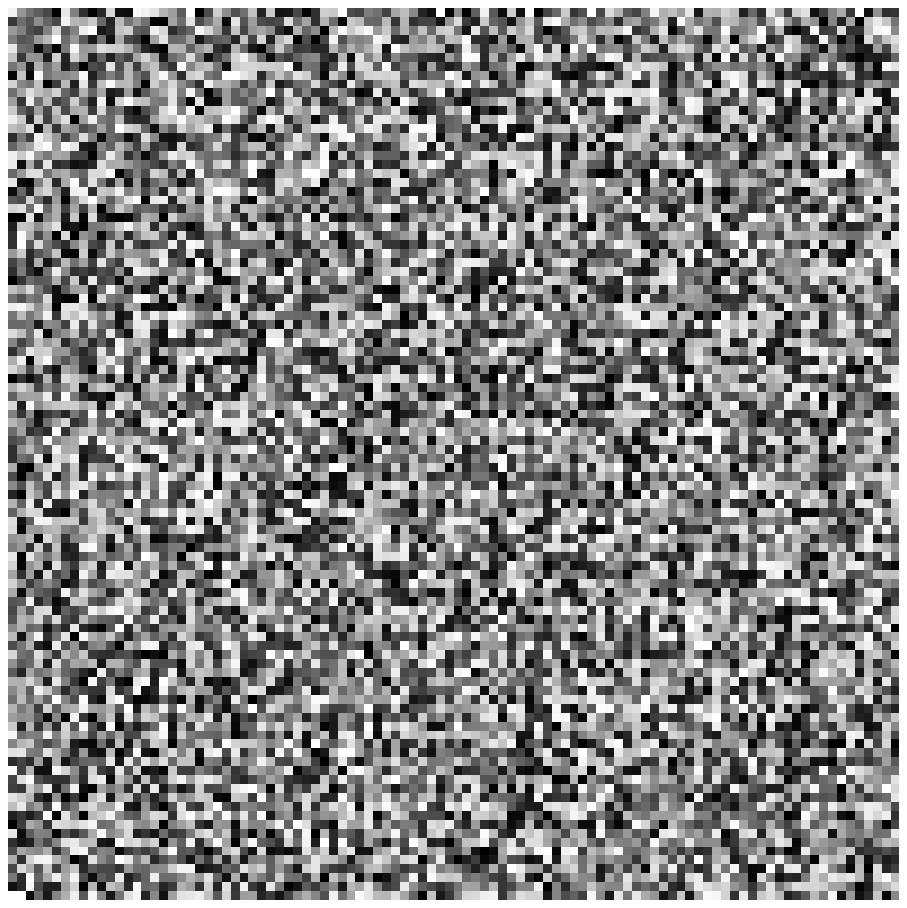}}}
  \subfigure[5250 steps]{%
  \label{fig:3b}
 \scalebox{0.40}{\includegraphics*[170,292][438,560]{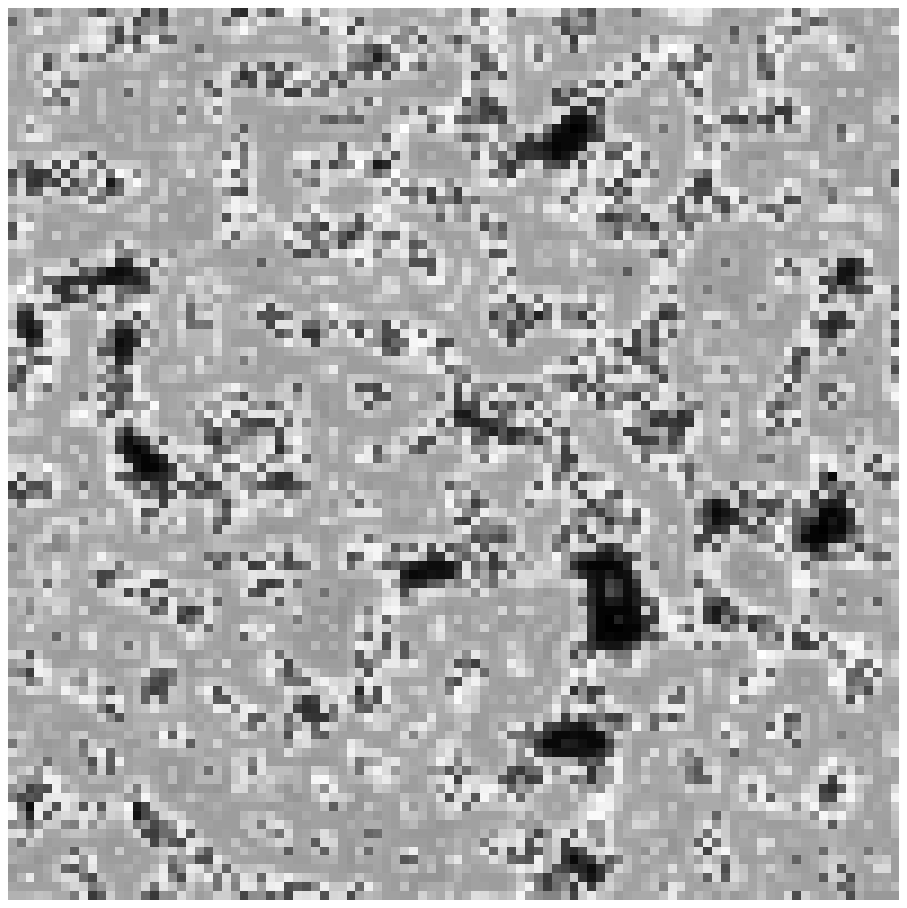}}}\\
   \subfigure[5300 steps]{%
  \label{fig:3c}
  \hspace{-0cm}\scalebox{0.40}{\includegraphics*[170,292][438,560]{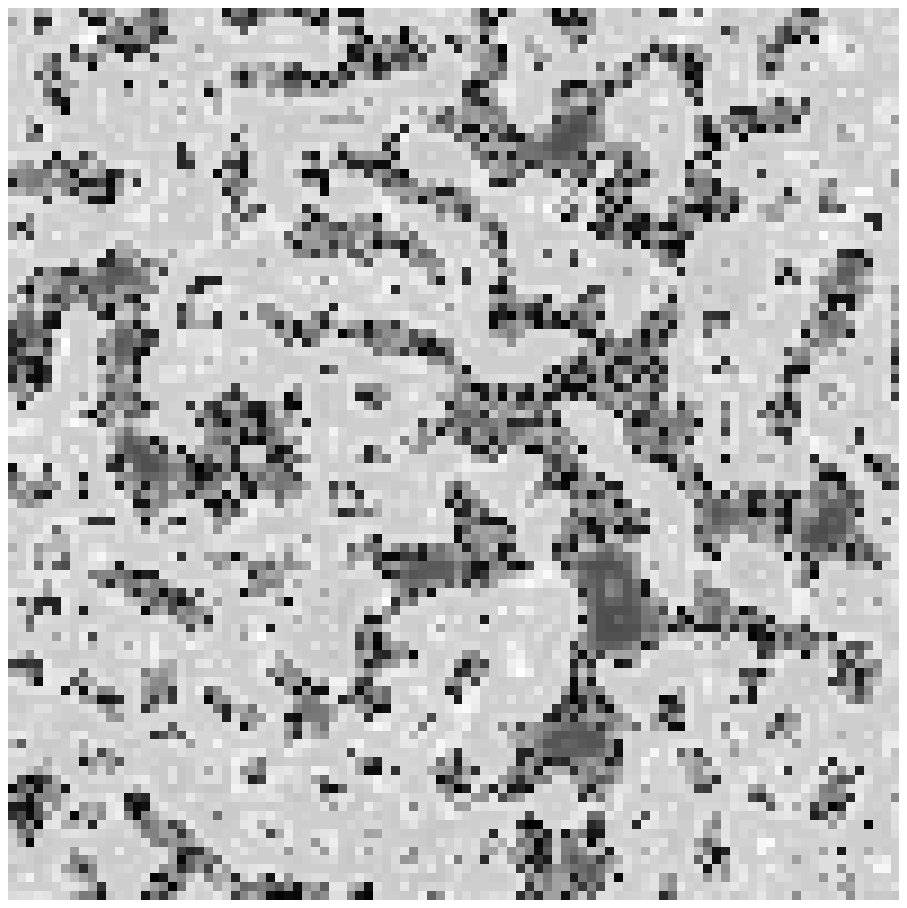}}}
  \subfigure[5350 steps]{%
  \label{fig:3d}
  \scalebox{0.40}{\includegraphics*[170,292][438,560]{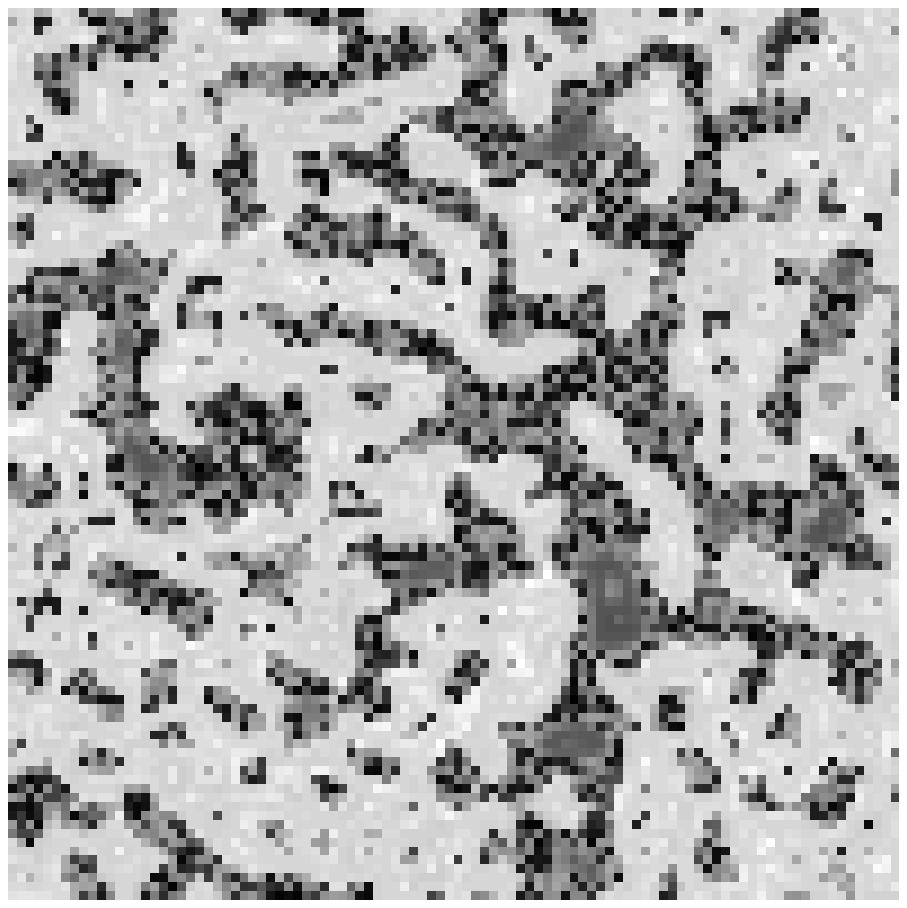}}}
  \caption{A typical simulation shows four snapshots of the evolution in
  two-dimensional space. The parameters is same as in Fig.~\ref{fig:1}, $\mu=0.02$, $\delta=35.84$,
  $\gamma=100$, $\beta_{0}=1800$ and $\varepsilon=0.38$. The figures plot
  susceptible density levels as a space on grey scale. The exposed and infected
  distribution has a qualitatively similar form.}\label{fig:3}
  \vspace{0.5cm}
\end{minipage}
\begin{minipage}[c]{0.45\textwidth}
   \subfigure[]{%
  \label{fig:4a}
 \scalebox{0.28}{\includegraphics*[90,262][480,560]{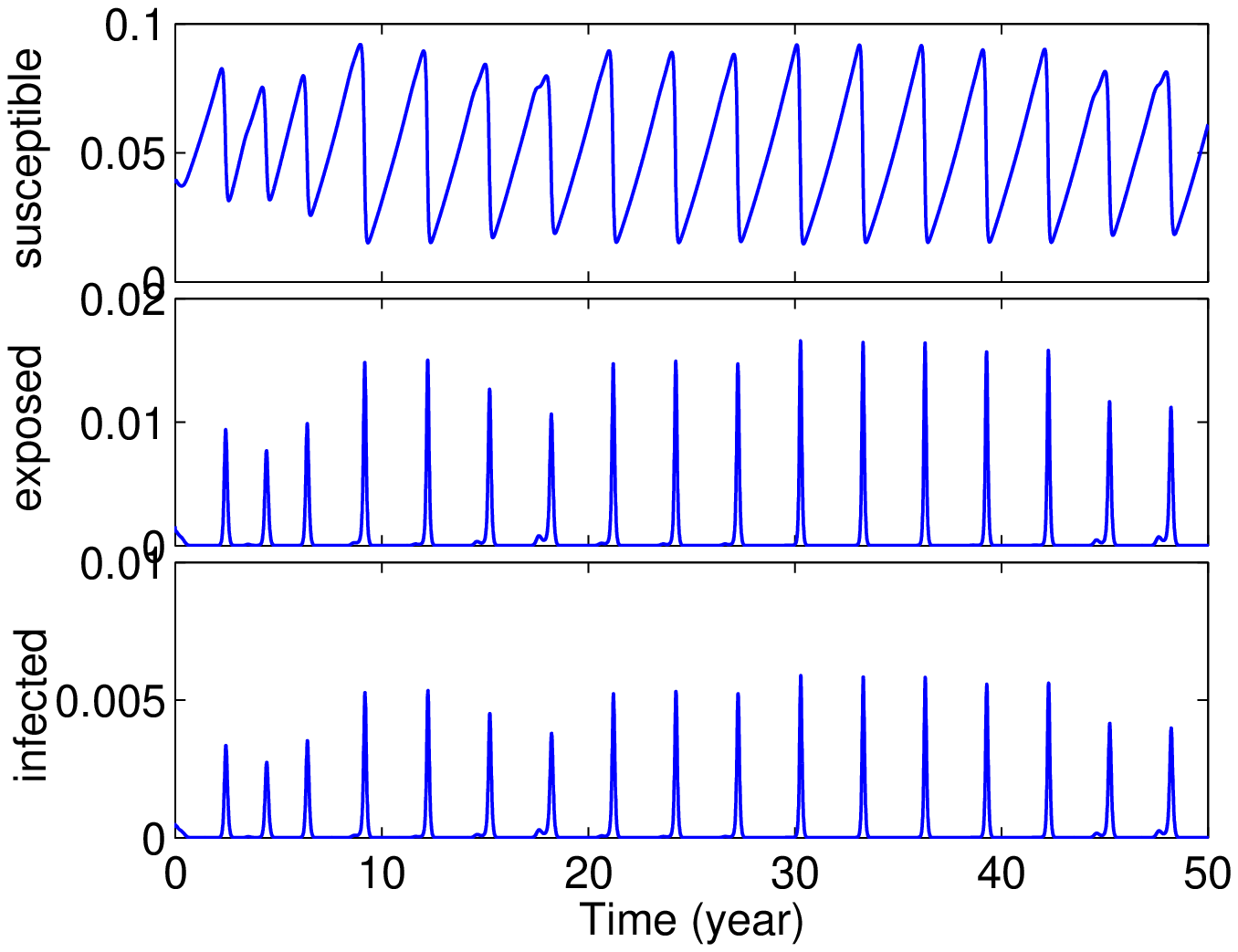}}}
  \subfigure[]{%
  \label{fig:4b}
 \scalebox{0.28}{\includegraphics*[90,262][480,560]{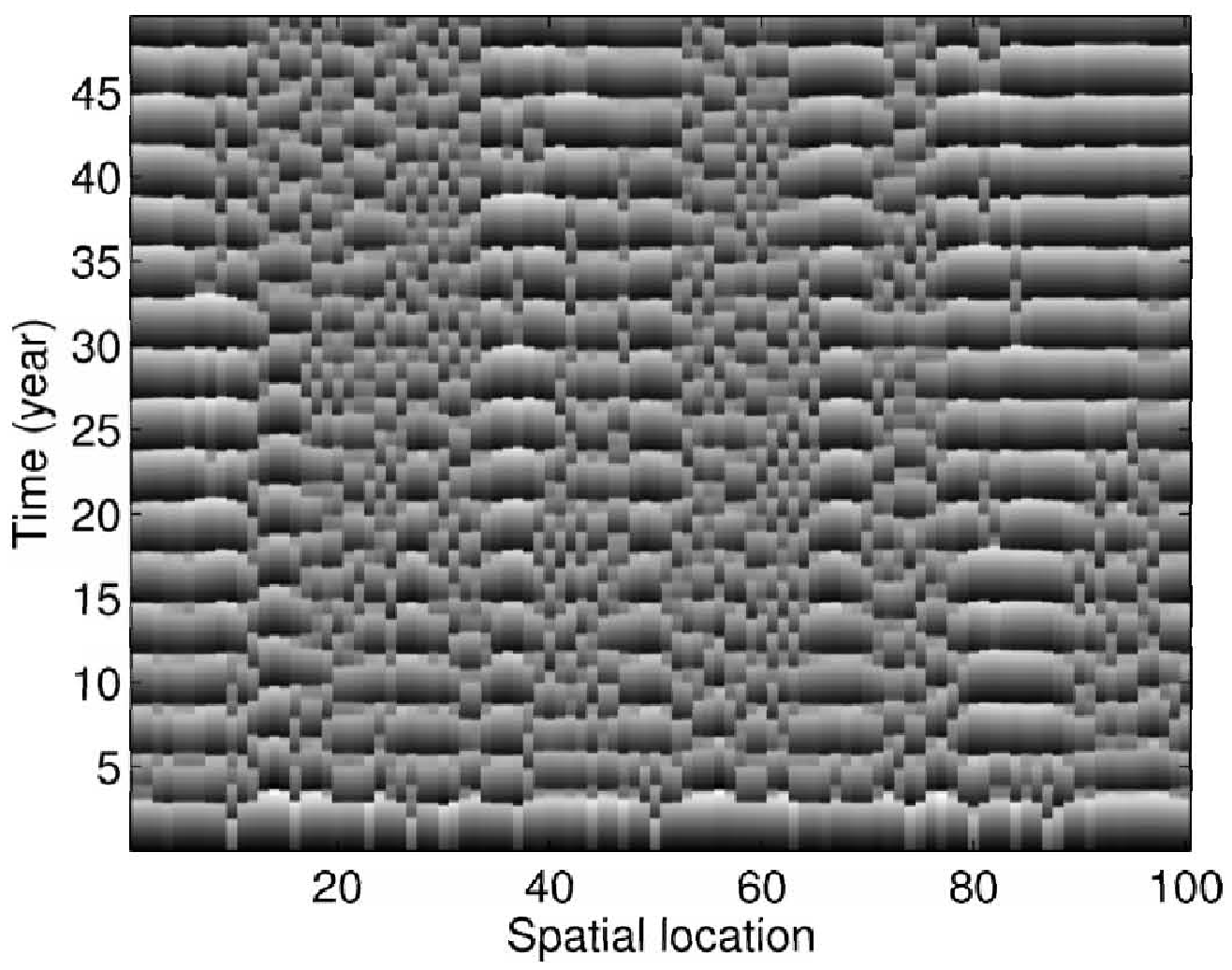}}}\\
  \subfigure[]{%
  \label{fig:4c}
\scalebox{0.28}{\includegraphics*[90,262][480,560]{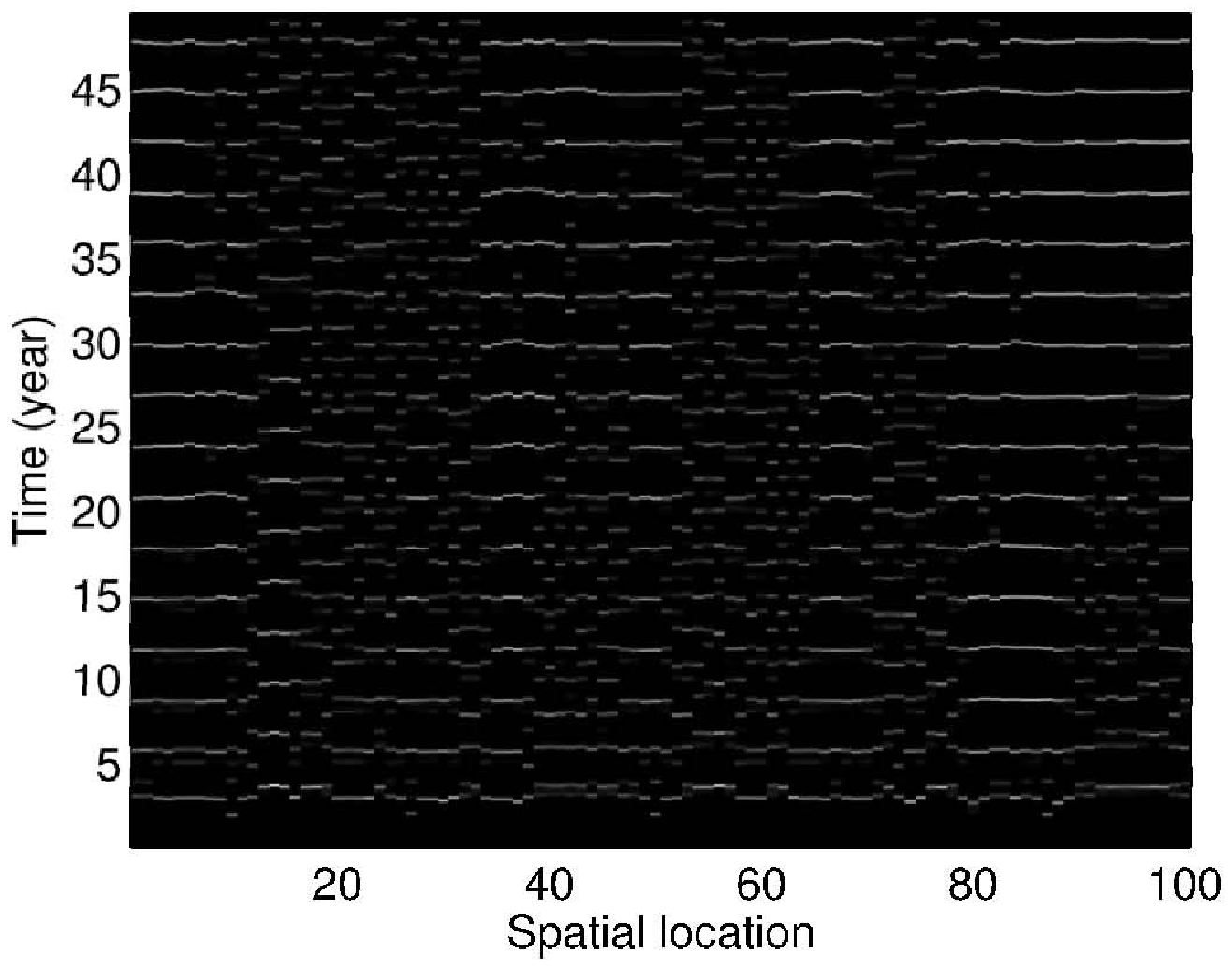}}}\hfill
  \subfigure[]{%
  \label{fig:4d}
  \includegraphics[width=3.5cm]{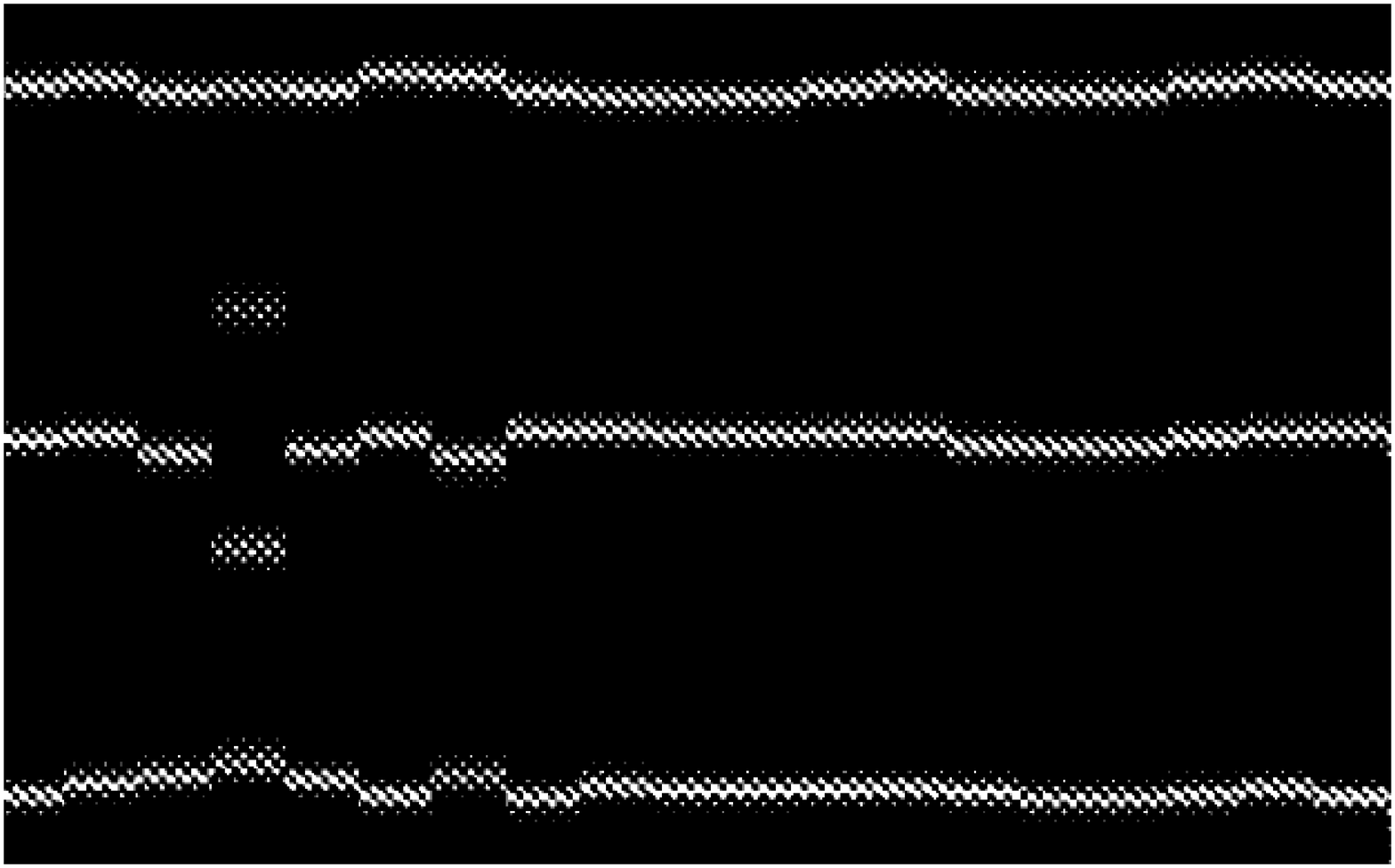}}
   \caption{The spatial period-2 structure results for the system \eqref{eq:3}
  in one-dimensional in cellular automata models and the parameters are the same as
  those in Fig.~\ref{fig:3}. We use a spatial domain of 100 lattices;
  10000 successive time iterations are plotted. In (a) we show  the species density as a
  function of time. The figure (b) and (c) show the susceptible and exposed
  density levels as a function of space and time on a grey scale respectively.
  The behavior of the infected is qualitatively similar. At this scale the spatial discretization in
  not really visible, and therefore we have enlarged one region of the plane in figure (c) as figure (d), in order
  to illustrate this discretization.}\label{fig:4}
\vspace{0.5cm}
\end{minipage}

 To furtherly investigate the impact of the fluctuating amplitude
on the dynamical patterns with fractal fronts (or spiral waves) in
two-dimensional and the spatial period-2 structure in
one-dimensional space respectively, we study the case when the SEIR
model is deeply in the domain of chaos and out of the domain of
chaos. The evolution of system \eqref{eq:3} is shown with
$\varepsilon=0.38$ in Fig.~\ref{fig:3} and Fig.~\ref{fig:4}. In
Fig.~\ref{fig:3b}-\ref{fig:3d} three snapshots are taken at 5250,
5300 and 5350 steps respectively. In Fig.~\ref{fig:3b} the rotating
spirals are not recognizable due to the irregular interfaces.
However, the spiral formation becomes visible when the interfacial
roughness grows by the infected invasion as demonstrated in
Fig.~\ref{fig:3d}. In Fig.~\ref{fig:3d} one can easily identify the
vortices and anti-vortices rotating clockwise and counterclockwise,
respectively. We have to emphasize that this pattern cannot be
characterized by a single length unit (e.g., correlation length)
because the main features of spirals (armlength, average curvature,
average distance, etc.) depend on the model parameters. But it may
be analysed by using some geometrical features method~\cite{Szab1}.
The armlength of these spiral waves are broad and do not easily
break, resulting in periodical recurrence of epidemic waves. The
spontaneous formation of spiral waves means the regularly recurrent
infection waves (Fig.~\ref{fig:3a}-\ref{fig:3d}). Similarly the
irregular spiral waves can also be observed even when the
fluctuating amplitude is much more than the critical value
$\varepsilon_{\rm c}$. These results are not shown in this paper.

\begin{minipage}[c]{0.45\textwidth}
   \subfigure[]{%
  \label{fig:5a}
 \scalebox{0.28}{\includegraphics*[90,262][480,560]{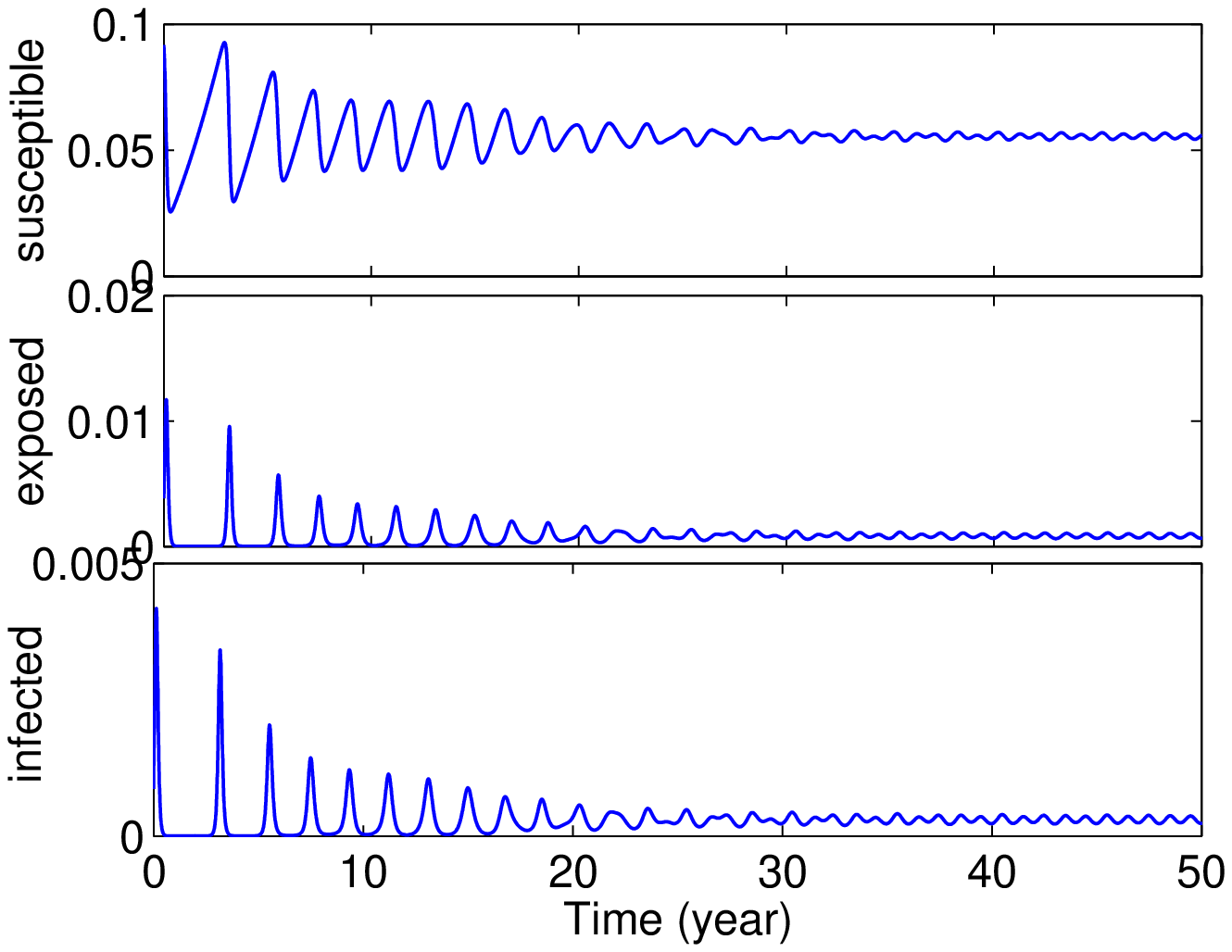}}}
  \subfigure[]{%
  \label{fig:5b}
 \scalebox{0.28}{\includegraphics*[90,262][480,560]{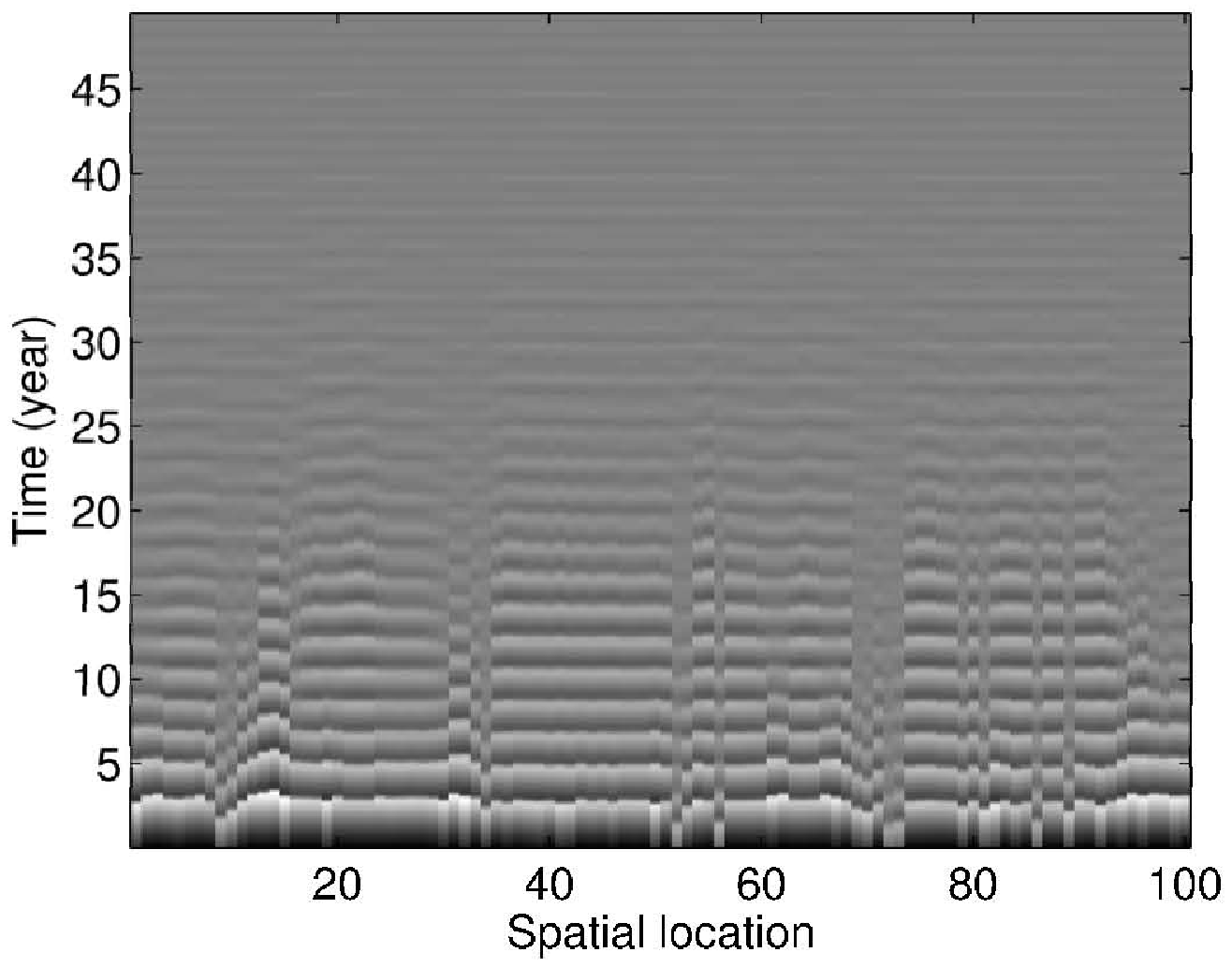}}}\\
 \caption{The spatio-temporal evolution of the system \eqref{eq:3}
  in one-dimensional in cellular automata models for the $\varepsilon<\varepsilon^{*}$ case,
   $\varepsilon=0.035$. In (a) we show  the species density as a function of time. The figure (b) is plot susceptible density as a function of space and time on a gray scale.
   The behavior of the exposed and infected is qualitatively similar.}\label{fig:5}
\vspace{0.5cm}
\end{minipage}

Figure~\ref{fig:5} shows the time evolution of the density of the
species and the spatio-temporal configurations of the system
\eqref{eq:3} at $\varepsilon=0.035$. The spatio-temporal evolution
of the expose and infected are similar with the susceptible's
(Figure~\ref{fig:5b}). It can be clearly noticed that spatial
period-2 structure disappear and the stationary state is a fixed
point with the decreasing of $\varepsilon$. The situation
corresponds to a low and persistent endemic infection in
Fig.~\ref{fig:5a}.  The oscillations decay to the fixed point when
the $\varepsilon$ is smaller than the critical value
$\varepsilon^{*}$ case.  The oscillations decay because these
infection clusters grow, the availability of infected hosts per
susceptible host is reduced, decreasing the number of new infectors.
In this case, the spontaneous formation of dynamical patterns is
qualitatively similar in Fig.~\ref{fig:3} in two-dimensional.

\section{Conclusions}

A realistic spatial epidemic with the individuals randomly moving in
its neighborhood has been modeled using cellular automata. Our
simulations demonstrate that the recurrent infectious waves exist
and persist in an exhaustive spatio-temporal. We have investigated
the dynamical patterns of system \eqref{eq:3} in one and two
dimensions perspectively. We show that the spiral waves recur
periodically and the recurrence is insensitive to the change of the
fluctuating amplitude $\varepsilon$ within the domain of chaos (the
fluctuation amplitude $\varepsilon>\varepsilon_{\rm c}$). Moreover
the dynamical patterns with fractal fronts grow stably. The system
also shows spatial period-2 structure in one dimension when
$\varepsilon$ is between $\varepsilon^{*}$ and $\varepsilon_{\rm c}$
outside the domain of chaos. It is interesting that  the spatial
period-2 structure will break and transform to spatial synchronous
configuration in the domain of chaos. Our results confirm that
populations embed and disperse more stably in space than they do in
non-spatial counterparts.

\section*{Acknowledgments}

This work was supported by the National Natural Science Foundation
of China under Grant No 10471040 and the Science Foundation of
Shan'xi Province No 2006011009.

\newpage 

\begin{thebibliography}{26}
\expandafter\ifx\csname
natexlab\endcsname\relax\def\natexlab#1{#1}\fi
\expandafter\ifx\csname bibnamefont\endcsname\relax
  \def\bibnamefont#1{#1}\fi
\expandafter\ifx\csname bibfnamefont\endcsname\relax
  \def\bibfnamefont#1{#1}\fi
\expandafter\ifx\csname citenamefont\endcsname\relax
  \def\citenamefont#1{#1}\fi
\expandafter\ifx\csname url\endcsname\relax
  \def\url#1{\texttt{#1}}\fi
\expandafter\ifx\csname urlprefix\endcsname\relax\def\urlprefix{URL
}\fi \providecommand{\bibinfo}[2]{#2}
\providecommand{\eprint}[2][]{\url{#2}}

\bibitem[{\citenamefont{Rohani and Miramontes}(1995{\natexlab{a}})}]{Rohani}
\bibinfo{author}{\bibfnamefont{P.}~\bibnamefont{Rohani}} \bibnamefont{and}
  \bibinfo{author}{\bibfnamefont{O.}~\bibnamefont{Miramontes}},
  \bibinfo{journal}{Proc. Roy. Soc. Lond B} \textbf{\bibinfo{volume}{260}},
  \bibinfo{pages}{335} (\bibinfo{year}{1995}{\natexlab{a}}).

\bibitem[{\citenamefont{Provata and Tsekouras}(2003)}]{Provata}
\bibinfo{author}{\bibfnamefont{A.}~\bibnamefont{Provata}} \bibnamefont{and}
  \bibinfo{author}{\bibfnamefont{G.~A.} \bibnamefont{Tsekouras}},
  \bibinfo{journal}{Phys. Rev. E} \textbf{\bibinfo{volume}{67}},
  \bibinfo{pages}{056602} (\bibinfo{year}{2003}).

\bibitem[{\citenamefont{Tsekouras and Provata}(2001)}]{Tsekouras}
\bibinfo{author}{\bibfnamefont{G.~A.} \bibnamefont{Tsekouras}}
  \bibnamefont{and} \bibinfo{author}{\bibfnamefont{A.}~\bibnamefont{Provata}},
  \bibinfo{journal}{Phys. Rev. E} \textbf{\bibinfo{volume}{65}},
  \bibinfo{pages}{016204} (\bibinfo{year}{2001}).

\bibitem[{\citenamefont{Provata et~al.}(1999)\citenamefont{Provata, Nicolis,
  and Baras}}]{Provata2}
\bibinfo{author}{\bibfnamefont{A.}~\bibnamefont{Provata}},
  \bibinfo{author}{\bibfnamefont{G.}~\bibnamefont{Nicolis}}, \bibnamefont{and}
  \bibinfo{author}{\bibfnamefont{F.}~\bibnamefont{Baras}}, \bibinfo{journal}{J.
  Chem. Phys.} \textbf{\bibinfo{volume}{110}}, \bibinfo{pages}{8361}
  (\bibinfo{year}{1999}).

\bibitem[{\citenamefont{Nicholas and Hogeweg}(1998)}]{Savill}
\bibinfo{author}{\bibfnamefont{J.~S.} \bibnamefont{Nicholas}} \bibnamefont{and}
  \bibinfo{author}{\bibfnamefont{P.}~\bibnamefont{Hogeweg}},
  \bibinfo{journal}{Proc Roy. Soc. Lond. B} \textbf{\bibinfo{volume}{265}},
  \bibinfo{pages}{25} (\bibinfo{year}{1998}).

\bibitem[{\citenamefont{Gurney et~al.}(1998)\citenamefont{Gurney, Veitch,
  Cruichshank, and Mcgeachin}}]{Gurney}
\bibinfo{author}{\bibfnamefont{W.~S.~C.} \bibnamefont{Gurney}},
  \bibinfo{author}{\bibfnamefont{A.~R.} \bibnamefont{Veitch}},
  \bibinfo{author}{\bibfnamefont{I.}~\bibnamefont{Cruichshank}},
  \bibnamefont{and}
  \bibinfo{author}{\bibfnamefont{G.}~\bibnamefont{Mcgeachin}},
  \bibinfo{journal}{Ecology} \textbf{\bibinfo{volume}{79}},
  \bibinfo{pages}{2516} (\bibinfo{year}{1998}).

\bibitem[{\citenamefont{Murray}(1993)}]{Murray}
\bibinfo{author}{\bibfnamefont{J.~D.} \bibnamefont{Murray}},
  \emph{\bibinfo{title}{Mathematical Biology}}
  (\bibinfo{publisher}{Springer-Verlag Berlin Heidelgerg},
  \bibinfo{year}{1993}).

\bibitem[{\citenamefont{Anderson and May}(1985)}]{Anderson}
\bibinfo{author}{\bibfnamefont{R.~M.} \bibnamefont{Anderson}} \bibnamefont{and}
  \bibinfo{author}{\bibfnamefont{R.~M.} \bibnamefont{May}},
  \bibinfo{journal}{Nature} \textbf{\bibinfo{volume}{318}},
  \bibinfo{pages}{323} (\bibinfo{year}{1985}).

\bibitem[{\citenamefont{Anderson and May}(1982)}]{Anderson2}
\bibinfo{author}{\bibfnamefont{R.~M.} \bibnamefont{Anderson}} \bibnamefont{and}
  \bibinfo{author}{\bibfnamefont{R.~M.} \bibnamefont{May}},
  \bibinfo{journal}{Science} \textbf{\bibinfo{volume}{215}},
  \bibinfo{pages}{1053} (\bibinfo{year}{1982}).

\bibitem[{\citenamefont{Dieckmann et~al.}(2000)\citenamefont{Dieckmann, Law,
  and Metz}}]{Dieckmann}
\bibinfo{author}{\bibfnamefont{U.}~\bibnamefont{Dieckmann}},
  \bibinfo{author}{\bibfnamefont{R.}~\bibnamefont{Law}}, \bibnamefont{and}
  \bibinfo{author}{\bibfnamefont{J.~A.~J.} \bibnamefont{Metz}},
  \emph{\bibinfo{title}{The Geometry of Ecological Interactions: Simplifying
  Spatial Complexity}} (\bibinfo{publisher}{Cambridge University Press},
  \bibinfo{address}{United Kingdom}, \bibinfo{year}{2000}).

\bibitem[{\citenamefont{Sherratt et~al.}(1997)\citenamefont{Sherratt, Eagan,
  and Lewis}}]{Sherratt}
\bibinfo{author}{\bibfnamefont{J.~A.} \bibnamefont{Sherratt}},
  \bibinfo{author}{\bibfnamefont{B.~T.} \bibnamefont{Eagan}}, \bibnamefont{and}
  \bibinfo{author}{\bibfnamefont{M.~A.} \bibnamefont{Lewis}},
  \bibinfo{journal}{Phil. Trans. Roy. Soc. Lond. B}
  \textbf{\bibinfo{volume}{352}}, \bibinfo{pages}{21} (\bibinfo{year}{1997}).

\bibitem[{\citenamefont{Weimar and Jean-Pierre}(1994)}]{Weimar}
\bibinfo{author}{\bibfnamefont{R.~J.} \bibnamefont{Weimar}} \bibnamefont{and}
  \bibinfo{author}{\bibfnamefont{B.}~\bibnamefont{Jean-Pierre}},
  \bibinfo{journal}{Phys. Rev. E} \textbf{\bibinfo{volume}{49}},
  \bibinfo{pages}{1749} (\bibinfo{year}{1994}).

\bibitem[{\citenamefont{Grenfell et~al.}(2001)\citenamefont{Grenfell, rnstad,
  and Kappey}}]{Grenfell_1}
\bibinfo{author}{\bibfnamefont{B.~T.} \bibnamefont{Grenfell}},
  \bibinfo{author}{\bibfnamefont{O.~N.} \bibnamefont{Bj{\o}rnstad}},
  \bibnamefont{and} \bibinfo{author}{\bibfnamefont{J.}~\bibnamefont{Kappey}},
  \bibinfo{journal}{Nature} \textbf{\bibinfo{volume}{414}},
  \bibinfo{pages}{716} (\bibinfo{year}{2001}).

\bibitem[{\citenamefont{Cummings et~al.}(2004)\citenamefont{Cummings, Huang,
  Endy, Nisalak, and K.~Ungchusak}}]{Cummings}
\bibinfo{author}{\bibfnamefont{D.~A.~T.} \bibnamefont{Cummings}},
  \bibinfo{author}{\bibfnamefont{N.~E.} \bibnamefont{Huang}},
  \bibinfo{author}{\bibfnamefont{T.~P.} \bibnamefont{Endy}},
  \bibinfo{author}{\bibfnamefont{A.}~\bibnamefont{Nisalak}}, \bibnamefont{and}
  \bibinfo{author}{\bibfnamefont{B.~D.~S.} \bibnamefont{K.~Ungchusak}},
  \bibinfo{journal}{Nature} \textbf{\bibinfo{volume}{427}},
  \bibinfo{pages}{344} (\bibinfo{year}{2004}).

\bibitem[{\citenamefont{Antal et~al.}(2001)\citenamefont{Antal, Droz, Lipowski,
  and \'{O}dor}}]{Antal}
\bibinfo{author}{\bibfnamefont{T.}~\bibnamefont{Antal}},
  \bibinfo{author}{\bibfnamefont{M.}~\bibnamefont{Droz}},
  \bibinfo{author}{\bibfnamefont{A.}~\bibnamefont{Lipowski}}, \bibnamefont{and}
  \bibinfo{author}{\bibfnamefont{G.}~\bibnamefont{\'{O}dor}},
  \bibinfo{journal}{Phys. Rev. E} \textbf{\bibinfo{volume}{64}},
  \bibinfo{pages}{036118} (\bibinfo{year}{2001}).

\bibitem[{\citenamefont{Droz and Pekalski}(2001)}]{Droz}
\bibinfo{author}{\bibfnamefont{M.}~\bibnamefont{Droz}} \bibnamefont{and}
  \bibinfo{author}{\bibfnamefont{A.}~\bibnamefont{Pekalski}},
  \bibinfo{journal}{Physica A} \textbf{\bibinfo{volume}{298}},
  \bibinfo{pages}{545} (\bibinfo{year}{2001}).

\bibitem[{\citenamefont{van Ballegooijen and Boerlijst}(2004)}]{Ballegooijen}
\bibinfo{author}{\bibfnamefont{W.~M.} \bibnamefont{van Ballegooijen}}
  \bibnamefont{and} \bibinfo{author}{\bibfnamefont{M.~C.}
  \bibnamefont{Boerlijst}}, \bibinfo{journal}{Proc. Nati. Acad. Sci.}
  \textbf{\bibinfo{volume}{101}}, \bibinfo{pages}{18246}
  (\bibinfo{year}{2004}).

\bibitem[{\citenamefont{Olsen and Schaffer}(1990)}]{Olsen}
\bibinfo{author}{\bibfnamefont{L.~F.} \bibnamefont{Olsen}} \bibnamefont{and}
  \bibinfo{author}{\bibfnamefont{W.~M.} \bibnamefont{Schaffer}},
  \bibinfo{journal}{Science} \textbf{\bibinfo{volume}{249}},
  \bibinfo{pages}{499} (\bibinfo{year}{1990}).

\bibitem[{\citenamefont{Grenfell et~al.}(1994)\citenamefont{Grenfell,
  Kleczkowski, Ellner, and Bolker}}]{Grenfell}
\bibinfo{author}{\bibfnamefont{B.~T.} \bibnamefont{Grenfell}},
  \bibinfo{author}{\bibfnamefont{A.}~\bibnamefont{Kleczkowski}},
  \bibinfo{author}{\bibfnamefont{S.~P.} \bibnamefont{Ellner}},
  \bibnamefont{and} \bibinfo{author}{\bibfnamefont{B.~M.}
  \bibnamefont{Bolker}}, \bibinfo{journal}{Phil. Trans. Roy. Soc. Lond. A}
  \textbf{\bibinfo{volume}{348}}, \bibinfo{pages}{515} (\bibinfo{year}{1994}).

\bibitem[{\citenamefont{Djebali}(2001)}]{Djebali}
\bibinfo{author}{\bibfnamefont{S.}~\bibnamefont{Djebali}},
  \bibinfo{journal}{Nonl. Anal.: Real World Applications}
  \textbf{\bibinfo{volume}{2}}, \bibinfo{pages}{417} (\bibinfo{year}{2001}).

\bibitem[{\citenamefont{Vecchio et~al.}(2006)\citenamefont{Vecchio, Primavera,
  and Carbone}}]{Vecchio}
\bibinfo{author}{\bibfnamefont{A.}~\bibnamefont{Vecchio}},
  \bibinfo{author}{\bibfnamefont{L.}~\bibnamefont{Primavera}},
  \bibnamefont{and} \bibinfo{author}{\bibfnamefont{V.}~\bibnamefont{Carbone}},
  \bibinfo{journal}{Phys. Rev. E} \textbf{\bibinfo{volume}{73}},
  \bibinfo{pages}{1913} (\bibinfo{year}{2006}).

\bibitem[{\citenamefont{Sherratt}(1996)}]{Sherratt2}
\bibinfo{author}{\bibfnamefont{J.~A.} \bibnamefont{Sherratt}},
  \bibinfo{journal}{Physica D} \textbf{\bibinfo{volume}{95}},
  \bibinfo{pages}{319} (\bibinfo{year}{1996}).

\bibitem[{\citenamefont{David et~al.}(2000)\citenamefont{David, Rohani,
  Benjamin, and Grenfell}}]{Earn}
\bibinfo{author}{\bibfnamefont{J.~D.~E.} \bibnamefont{David}},
  \bibinfo{author}{\bibfnamefont{P.}~\bibnamefont{Rohani}},
  \bibinfo{author}{\bibfnamefont{M.~B.} \bibnamefont{Benjamin}},
  \bibnamefont{and} \bibinfo{author}{\bibfnamefont{B.~T.}
  \bibnamefont{Grenfell}}, \bibinfo{journal}{Science}
  \textbf{\bibinfo{volume}{287}}, \bibinfo{pages}{667} (\bibinfo{year}{2000}).

\bibitem[{\citenamefont{Hassell et~al.}(1991)\citenamefont{Hassell, Comins, and
  May}}]{Hassell1}
\bibinfo{author}{\bibfnamefont{M.~P.} \bibnamefont{Hassell}},
  \bibinfo{author}{\bibfnamefont{H.~N.} \bibnamefont{Comins}},
  \bibnamefont{and} \bibinfo{author}{\bibfnamefont{R.~M.} \bibnamefont{May}},
  \bibinfo{journal}{Nature} \textbf{\bibinfo{volume}{353}},
  \bibinfo{pages}{255} (\bibinfo{year}{1991}).

\bibitem[{\citenamefont{Rohani and Miramontes}(1995{\natexlab{b}})}]{Rohani2}
\bibinfo{author}{\bibfnamefont{P.}~\bibnamefont{Rohani}} \bibnamefont{and}
  \bibinfo{author}{\bibfnamefont{O.}~\bibnamefont{Miramontes}},
  \bibinfo{journal}{Proc. Roy. Soc. Lond. B} \textbf{\bibinfo{volume}{260}},
  \bibinfo{pages}{335} (\bibinfo{year}{1995}{\natexlab{b}}).

\bibitem[{\citenamefont{Szab\'{o} and Szolnoki}(2002)}]{Szab1}
\bibinfo{author}{\bibfnamefont{G.}~\bibnamefont{Szab\'{o}}} \bibnamefont{and}
  \bibinfo{author}{\bibfnamefont{A.}~\bibnamefont{Szolnoki}},
  \bibinfo{journal}{Phys. Rev. E} \textbf{\bibinfo{volume}{65}},
  \bibinfo{pages}{036115} (\bibinfo{year}{2002}).

\end{thebibliography}

\end{document}